\documentclass{osa-article}

\journal{oe}

\usepackage{color}
\usepackage{amsmath,amssymb}
\usepackage[version=4]{mhchem}
\begin{document}

\title{Optical Frequency Combs  from High-Order Sideband Generation}

\author{Darren C. Valovcin$^{1, 2}$, Hunter B. Banks$^{3}$, Shawn Mack$^4$, Arthur C. Gossard$^5$, Kenneth West $^6$, Loren Pfeiffer$^6$ and Mark S. Sherwin$^{1, 2,*}$}

\address{$^1$Physics Department, University of California, Santa Barbara, USA\\
$^2$Institute for Terahertz Science and Technology, University of California, Santa Barbara, USA\\
$^3$Washington University School of Medicine, Department of Radiology, St. Louis, Missouri, USA\\
$^4$U.S. Naval Research Laboratory, Washington, DC, USA\\
$^5$Materials Department, University of California, Santa Barbara, USA\\
$^6$Electrical Engineering Department, Princeton University, Princeton, NJ, USA}

\email{$^*$sherwin@physics.ucsb.edu} 

\begin{abstract}
We report on the generation of frequency combs from the recently-discovered phenomenon of high-order sideband generation (HSG).  A near-band gap continuous-wave (cw) laser with frequency $f_\text{NIR}$ was transmitted through an epitaxial layer containing GaAs/AlGaAs quantum wells that were driven by quasi-cw in-plane electric fields $F_\text{THz}$ between 4 and 50 kV/cm oscillating at frequencies $f_\text{THz}$ between 240 and 640 GHz. Frequency combs with teeth at $f_\text{sideband}=f_\text{NIR}+nf_\text{THz}$ ($n$ even) were produced, with maximum reported $n>120$, corresponding to a maximum comb span $>80$ THz.  Comb spectra with the identical product $f_\text{THz}\times F_\text{THz}$ were found to have similar spans and shapes in most cases, as expected from the picture of HSG as a scattering-limited electron-hole recollision phenomenon. The HSG combs were used to measure the frequency and linewidth of our THz source as a demonstration of potential applications.
\end{abstract}

\bibliographystyle{unsrt}

\section{Introduction}
In recent years, optical frequency combs (OFCs) have been used for a wide variety of applications~\cite{Kim2016MLLReview,Newbury2011} such as calibration of astronomical spectrograms~\cite{Steinmetz2008}, high-resolution metrology~\cite{Udem2002}, and optical communications~\cite{Yi2010}. Three primary mechanisms for generating OFCs are widely used at this time. Mode-locked lasers (MLLs) can generate OFCs~\cite{Kim2016MLLReview, Quinlan2009MLLReview} with bandwidths that exceed one octave~\cite{Diddams2000MLLWidth}. The spacing of the teeth in such OFCs is determined by the length of the MLL cavity.  This tooth spacing is limited to several GHz, and can be tuned only by changing the length of the MLL cavity~\cite{TC2013-Review, Quinlan2009MLLReview}. Continuous-wave (CW) lasers coupled to micro-resonators can generate OFCs called microcombs through cascaded four-wave mixing processes~\cite{DelHaye2007, PASQUAZI2018}. The spacing of microcomb teeth is determined by the dimensions of the micro-resonator.  This tooth spacing can exceed 1 THz, but is tunable only to the limited extent that the dimensions of the solid microresonator can be changed  \cite{PASQUAZI2018}. The simplest and oldest method of generating OFCs is electro-optic modulation (EOM) of a CW laser~\cite{TC2013-Review, Murata2000}. The spacing of the teeth in such OFCs is determined by the radio-frequency (RF) (typically at 10s of GHz) used to drive the modulator. Since the CW laser and the RF source can be varied continuously and independently, both the tooth spacing and comb center are widely variable. However, the typical bandwidth of OFCs based on a single EOM is generally smaller than 1 THz~\cite{TC2013-Review}. Octave-spanning OFCs have been generated by injecting the output of an OFC based on EOMs into a series of optical amplifiers, highly nonlinear fibers, filters, and other components~\cite{Beha:17}.

High-order sideband generation (HSG) is a recently-discovered phenomenon that generates OFCs~\cite{Zaks2012, bulkGaAs, banksAntenna,banksPRLPhonons}. As in combs generated with EOMs, the center frequency and tooth spacing of combs based on HSG are both continuously and widely tunable.  However, the tooth spacings associated with combs based on HSG are much larger, ranging from of 100s of GHz to more than 1 THz. In HSG, a weak near-infrared (NIR) laser tuned near resonance with the exciton energy in a semiconductor is incident on a sample that is driven by a strong THz field. Similar to the atomic phenomenon of high-order harmonic generation (HHG)~\cite{MacklinHHG}, a three-step model~\cite{corkum3step} can be used to describe the following dynamics~\cite{rblPrediction, DBR}; the NIR laser creates an electron and hole in the semiconductor which the THz field (i) ionizes, (ii) accelerates away from and then back towards each other and (iii) causes to recollide, emitting a sideband photon with higher energy than the NIR laser. Recent time-resolved experiments have observed a delay between electron-hole pair creation and sideband emission~\cite{Langer2016}, which supports this recollision model in solids. 

In this paper, we demonstrate HSG-based combs with tooth spacing ranging from 0.494 to 1.266 THz created by THz electric fields ranging from 7.3 to 54 kV/cm.  We find that comb spectra with the identical product $f_\text{THz}\times F_\text{THz}$ have similar spans and shapes in most cases, as expected from a picture in which sidebands are attenuated by scattering processes with characteristic time scales that are independent of the THz frequency. As an example application, we use the OFCs produced by HSG to characterize the linewidth and frequency of our THz source, the UCSB mm-wave Free Electron Laser (FEL). The electric fields required for HSG are achievable over small areas with modern sub-THz sources and careful microwave engineering, which would enable the widespread deployment of OFCs based on HSG~\cite{Zaks2012}.      
\section{Experimental Setup}
\begin{figure}
    \centering
    \includegraphics{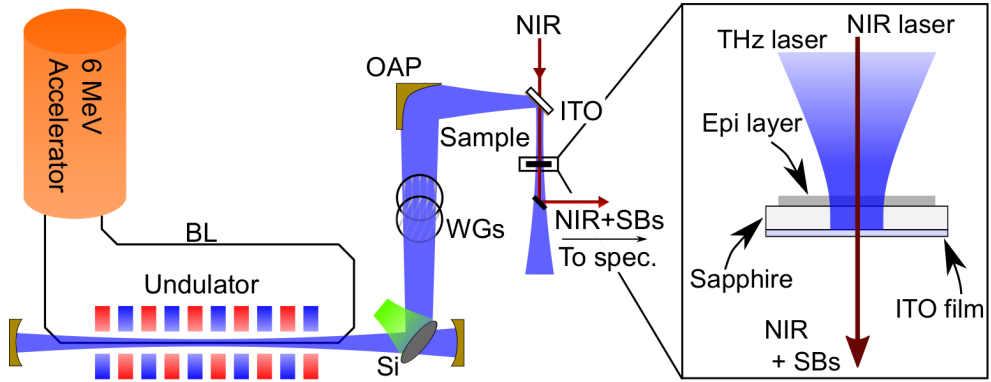}
    \caption{Schematic of the experimental setup. (Left) The Free Electron Laser (FEL) consists of the  6 MV linear electrostatic accelerator, electron beam line (BL), undulator (alternating red and blue rectangles) and silicon cavity dump out-coupler (Si). THz power is controlled by crossing two wiregrid polarizers (WGs). It is focused with an off-axis parabolic mirror (OAP) and an indium-tin oxide (ITO) coated slide combines the THz and near-ir (NIR) radiation. The generated sidebands (SBs) are sent into an imaging spectrometer (spec.). (Right) Enlarged view of sample. NIR and THz lasers are colinearly incident on the sample. The sample consists of an epitaxial (epi) layer containing GaAs/AlGaAs quantum wells transferred onto sapphire backed with an ITO coating. The ITO coating reflects the THz radiation to enhance the field in the sample while transmitting the SBs and NIR radiation.}
    \label{fig:1expSetup}
\end{figure}

Experiments were performed using the UCSB mm-wave Free Electron Laser (FEL) as a source of tunable, high-power, spectrally bright THz radiation, see Fig.~\ref{fig:1expSetup}. The UCSB FEL is capable of producing radiation from 240 GHz to 4.5 THz. Out-coupling of the THz radiation from the FEL in these experiments was performed using the cavity dump coupler~\cite{susumuCDC}, which allows for peak powers of up to 100 kW in 40 ns pulses. The THz power was controlled by two wire grid polarizers; the first was free to rotate while the second was kept fixed to ensure the polarization was constant, and the power was measured using a Thomas-Keating absolute power meter. An off-axis parabolic mirror focused the radiation to a diffraction limited spot. To control the relative NIR and THz alignment, an indium tin oxide (ITO) coated glass slide reflected the THz onto the sample while transmitting the NIR.

Two samples were used in this experiment. The first contained 10$\times$ 10 nm \ce{GaAs}/\ce{Al_{0.3}Ga_{0.7}As} quantum wells (QWs) which had been epitaxially transferred to a sapphire substrate for support. A thin film of ITO was deposited on the backside of the sapphire to create a strong THz cavity to enhance the field in the QWs, see Fig.~\ref{fig:1expSetup} and \cite{DBR} for details. The second sample contained 20$\times$ 5 nm \ce{GaAs}/\ce{Al_{0.3}Ga_{0.7}As} and was processed in the same fashion. The samples were mounted in a closed-cycle cryostat and held at 15 K for all experiments. The NIR excitation wavelength was kept at 798.3 nm for the 10nm QWs, and 764.1 nm for the 5nm QWs, which correspond to the heavy-hole exciton resonance in each sample. The NIR power was fixed at 50 mW ($\sim250$ V/cm), and the polarization was fixed at 90$^\circ$ relative to the THz polarization to minimize effects due to dynamical birefringence~\cite{DBR}. Reported THz field strengths are those in the QWs, including dielectric screening, and have an error of  $\pm 1$ kV/cm. 

When the NIR and THz simultaneously excited the sample, sidebands were generated at $f_\text{SB} = f_\text{NIR} + nf_\text{THz}$ (n even). Low-order sidebands, $n\lesssim8$, and negative-order sidebands were detected with a double grating monochromator and PMT. In order to calibrate the power in the sidebands, the (heavily-attenuated) NIR laser was measured using the same monochromator and PMT. Higher-order positive sidebands were measured in parallel with an imaging spectrometer and electron-multiplied CCD camera (EMCCD)~\cite{DBR}. 

\section{Optical frequency combs from HSG}
\begin{figure}
    \centering
    \includegraphics{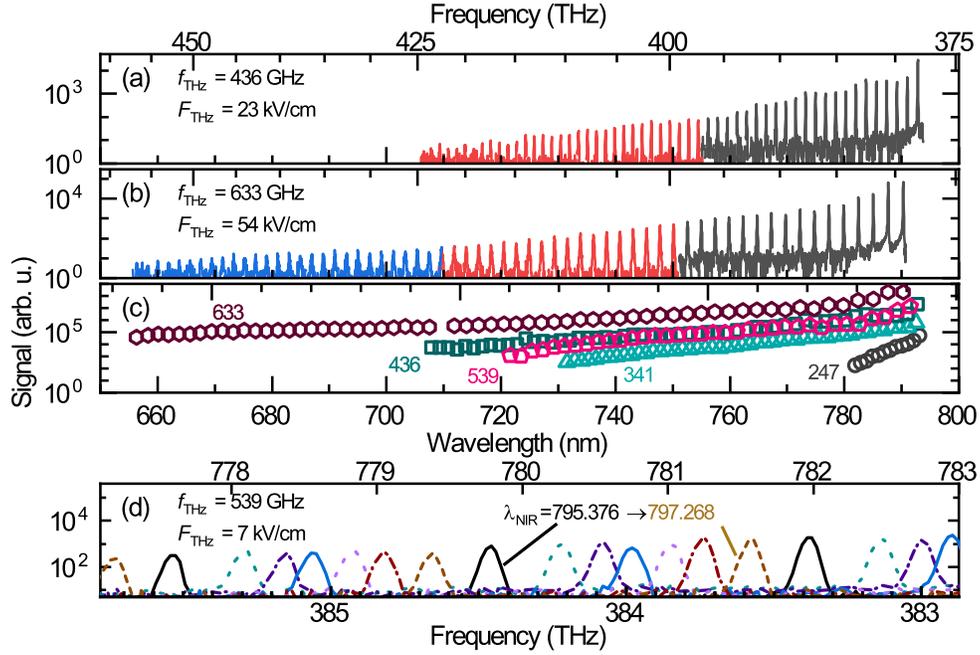}
    \caption{Optical frequency combs with variable tooth spacing (a-c) and offset frequency (d) from HSG in 10 nm GaAs quantum wells. (a,b) Spectra of positive orders measured by the CCD for (a) $f_\text{THz}$=436 GHz and $F_\text{THz}$=23 kV/cm and (b) $f_\text{THz}$=633 GHz and $F_\text{THz}$=54 kV/cm. The different colors correspond to spectra taken with different grating positions in the monochromator, and integration times ranging from $\sim$25 to $\sim$250 FEL pulses. (c) Strength of each sideband (area under Gaussian fit) vs. wavelength for HSG spectra generated by THz fields of 7 kV/cm at 247 GHz (black circles), 24 kV/cm at 341 GHz (turquoise triangles), 23 kV/cm at 436 GHz (green squares), 48 kV/cm at 539 GHz (pink pentagons), and 54 kV/cm at 633 GHz (brown hexagons). Spectra are offset slightly for visibility. For each comb, the tooth spacing is $2 \times f_\text{THz}$.  (d) Spectra measured by the CCD for a THz field of 7 kV/cm at 539 GHz with several NIR excitation wavelengths spaced by roughly 0.3 nm. Values of $\lambda_\text{NIR}$, in nm, are indicated for the smallest (black) and largest (gold) NIR laser excitation wavelengths used.}
    \label{fig:2rawcombs}
\end{figure}

Examples of OFCs generated using various THz field strengths and frequencies are shown in Fig.~\ref{fig:2rawcombs}(a-c). The spectrometer bandwidth was approximately 50 nm for a single grating orientation, and up to three images were taken at different grating orientations to cover the full comb bandwidth. In each image, the integration time was adjusted to maximize the dynamic range while preventing the camera from saturating. Each spectrum was normalized to the number of FEL pulses during the collection, so spectra from each image are directly comparable. An OFC spanning roughly 50 THz, generated by a 23 kV/cm THz field at 436 GHz and collected using two grating orientations, is shown in Fig.~\ref{fig:2rawcombs}(a). A much broader OFC spanning more than 80 THz, generated by a 54 kV/cm field at 633 GHz and collected using three grating orientations, is shown in Fig.~\ref{fig:2rawcombs}(c).

In order to compare the strengths of sidebands from different field strength and frequency combinations, each sideband was fit to a Gaussian, and the strength of each sideband was defined as the area under the Gaussian. Figure~\ref{fig:2rawcombs}(c) shows sideband strengths for the two combs shown in Fig.~\ref{fig:2rawcombs}(a,b), as well as combs generated using three additional THz frequencies. Each frequency was chosen to match a resonance of the sample-sapphire-ITO cavity~\cite{DBR}, while the field strengths (7 kV/cm at 247 GHz, 24 kV/cm at 341 GHz, and 48 kV/cm at 539 GHz) were the maximum fields in the sample achievable from the THz source. While each spectrum in Fig.~\ref{fig:2rawcombs}(c) displays a roughly exponential decay with increasing order, the rate of decay for each spectrum varies drastically. At 247 GHz, the low frequency and field strength resulted in a rapid fall-off in sideband strength with order compared to the higher driving frequencies, in part because of the relatively low field strength available at that frequency.

In addition to continuously tuning the comb spacing by changing the THz frequency, it was also possible to continuously tune the comb offset frequency by changing the NIR laser excitation frequency. Figure~\ref{fig:2rawcombs}(d) shows an example of different OFCs produced with identical THz fields of 7 kV/cm and 539 GHz, but various NIR frequencies. When exciting near 800nm and driving at 539 GHz, the comb spacing is approximately 2.3 nm so by tuning the excitation frequency through 2 nm around the heavy hole exciton resonance, it was possible to achieve arbitrary comb offset frequencies. 

\section{Controlled Manipulation of Comb Bandwidths}
    
As can be seen from Fig.~\ref{fig:2rawcombs}(c), the sidebands produced at various THz field strengths and frequencies can produce a wide variety of comb bandwidths. In high-harmonic generation, the three step model predicts a maximum harmonic energy, $E_\text{max}$, which scales with the strong laser's electric field, $F$, and frequency, $\omega$, as $E_\text{max} = I_p + 3.17(e^2F^2)/(4m\omega^2)$, where $I_p$ is the ionization potential of the studied atom, $e$ is the electron charge, and $m$ the electron mass~\cite{corkum3step, lewenstein3step}. This simple cutoff relation has been observed in experiments~\cite{Krause3Up}. For HSG, such cutoffs are predicted by theory in the absence of scattering~\cite{rblPrediction, 2008YanTheory,rblIntoGap}, but have not been observed in experiment~\cite{Zaks2012, DBR, banksPRLPhonons}. In the presence of scattering, electrons and holes may be scattered before they recollide. If they do scatter, they are unlikely to produce a sideband. Because the time between electron-hole excitation and recollision increases with increasing order, the attenuation of sidebands due to scattering increases with increasing order, making it difficult to observe the predicted plateau and cutoff. For the spectra shown in Fig.~\ref{fig:2rawcombs}, the predicted cutoff wavelength is nearly 520 nm at 247 GHz or 150nm at 633 GHz, much higher than the observed comb bandwidths. One should not expect to observe sidebands associated with recollisions that take much longer than than typical scattering times, which are $\sim200$ fs in these samples~\cite{banksPRLPhonons}.

\begin{figure}
    \centering
    \includegraphics{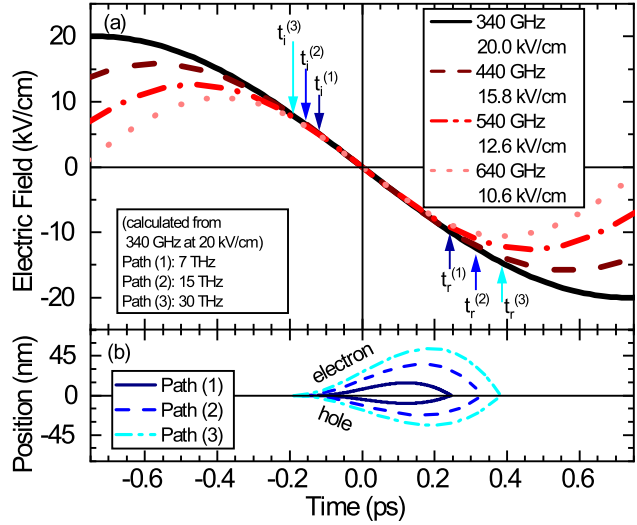}
    \caption{Classical calculations of carrier dynamics. (a) Time-domain waveforms of the THz field calculated for four THz driving waves chosen such that $f_\text{THz}\times F_\text{THz}=6.8\text{ THz} \cdot$ kV/cm. The ionization and recollision times are $t_i^{(m)}$ and $t_r^{(m)}$, respectively, for sidebands offset from the NIR laser by 7 THz ($m=1$), 15 THz ($m=2$), and 30 THz ($m=3$), as calculated from the semiclassical three-step model assuming parabolic bands with an electron with an effective mass $0.063m_e$ and a hole with an effective mass $0.1m_e$~\cite{banksPRLPhonons} ($m_e$ is the free electron mass) driven by a 340 GHz, 20 kV/cm field. Each of the three steps occur during a portion of the driving field in which the electric field is approximately linear in time. (b) Classical trajectories of the electron and hole responsible for generating the sidebands offset from the NIR laser by (1) 7 THz (2) 15 THz and (3) 30 THz. Electrons and holes can travel several 10s of nanometers through the crystal in this recollisions process.}
    \label{fig:4fieldschematic}
\end{figure}
If scattering dominates the rate of decay of sidebands with increasing order, one might expect that, for different combinations of driving frequency and THz field, the sidebands associated with trajectories of equal duration should have roughly equal strengths. Motivated by this notion, we describe a simple scaling law which relates overall comb bandwidths to the driving field strength and frequency. Figure \ref{fig:4fieldschematic}(a) depicts a portion of a period of the THz waveform at four different field strength and frequency combinations chosen to give identical electric fields, $F_\text{THz}(t)$ near the zero-crossing of the THz field. The dynamics from the three-step model~\cite{corkum3step} --assuming simple parabolic bands with conduction band effective mass $0.063m_e$ and a valence band effective mass $0.1m_e$ ($m_e$ is the free electron mass)~\cite{banksPRLPhonons} can be solved numerically to find the recollision times and trajectories (see Fig~\ref{fig:4fieldschematic}(b)) for the injection times associated with various sideband offset frequencies at these fields. The three arrows in Fig.~\ref{fig:4fieldschematic}(a) show the ionization ($t_i^{(m)}$) and recollision ($t_r^{(m)}$) times for three different sideband offset frequencies, as calculated for the 20 kV/cm field at 340 GHz. From Fig.~\ref{fig:4fieldschematic}(a), we can see that the majority of the sidebands that are created in Fig.~\ref{fig:2rawcombs} at a driving frequency of 340 GHz are created during the linear portion of the driving field where

\begin{equation}
    F_\text{THz}(t) = F_\text{THz}\sin(2\pi f_\text{THz} t) \approx  F_\text{THz} 2\pi f_\text{THz} t \label{eq:linearapprox}
\end{equation}
with $F_\text{THz}$ and $f_\text{THz}$ being the electric field's amplitude and frequency, respectively.

For THz fields satisfying $F_\text{THz} f_\text{THz}\approx\text{constant}$, if an electron and hole are injected and recollide while the THz field is close to linear in time, they should produce sidebands with roughly the same strength for different THz frequencies. If typical scattering times are sufficiently short, one also expects similar comb bandwidths. Since the driving frequency is changing, however, the spacing between comb teeth will vary.

\begin{figure}
    \centering
    \includegraphics{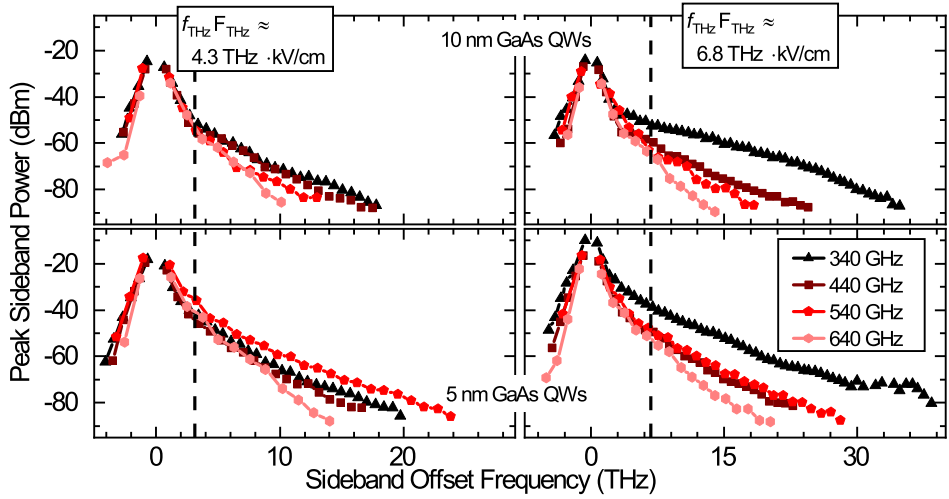}
    \caption{Peak sideband power for (top) a 10 nm GaAs QW structure and (bottom) a 10 nm AlGaAs QW sample when the driving THz is tuned such that $f_\text{THz}F_\text{THz}\approx4.3$ THz$\cdot$kV/cm (left) and 6.8 THz$\cdot$kV/cm (right). Dashed lines indicate the offset frequency of a sideband emitted at time when the 640 GHz field profile deviates from the linear approximation by 10\%; 3 THz at low field and 7 THz at high field. (Left) Field strengths are 13 kV/cm at 340 GHz, 10 kV/cm at 440 GHz, 8 kV/cm at 540 GHz and 7 kV/cm at 640 GHz. (Right) Field strengths are 20 kV/cm at 340 GHz, 16 kV/cm at 440 GHz, 13 kV/cm at 540 GHz and 11 kV/cm at 640 GHz.}
    \label{fig:5bandwidthComps}
\end{figure}

To test this scaling law, we drove two quantum well (QW) samples with several combinations of frequency and field with different products $F_\text{THz} f_\text{THz}$. The resulting peak sideband powers (powers during the 40 ns FEL pulse, in dBm) are shown in Fig.~\ref{fig:5bandwidthComps}. The top two plots show the results from a clean 10 nm GaAs QW with a heavy-hole exciton (HHX) linewidth of of approximately 2.0 meV, while the bottom plots show the results from a 5 nm QW in which the well-width fluctuations result in a proportionally broadened HHX linewidth of 6.3 meV (see \cite{DBR} for details on samples). For each quantum well studied, the left (right) plots show the peak sideband powers for $f_\text{THz}F_\text{THz}$=4.3 (6.8) THz$\cdot$kV/cm. 

In the 10nm QW sample at lower electric fields (Fig.~\ref{fig:5bandwidthComps} upper left), each frequency and field combination resulted in equal sideband powers up to around a 3 THz sideband offset frequency. Above a 3 THz offset, the 340 GHz comb and 440 GHz combs resulted in higher sideband powers than the 540 and 640 GHz combs. At 340 and 440 GHz, comb widths of 18 THz were observed, decreasing to 14 THz and 10 THz at 540 and 640 GHz driving frequency, respectively. Generally, the lower frequency driving fields resulted in combs with larger bandwidth and higher sideband power than higher frequency driving fields. 

To understand why each comb produced slightly different sideband strengths and comb bandwidths, counter to the initial hypothesis, we consider where the linear electric field approximation begins to break down. Figure \ref{fig:4fieldschematic} shows that higher frequencies deviate from the linear approximation more quickly than lower frequencies. For 7 kV/cm at 640 GHz, the time $t_r$ at which the field deviates from the linear approximation by 10\% after the zero crossing is associated with the recollision of an electron-hole pair responsible for the emission of a 3 THz sideband (depicted by a vertical dashed line in the left side of Fig.~\ref{fig:5bandwidthComps}). Below this sideband offset, all combs for the low-field driven 10nm GaAs sample agreed quite well. Above this 3 THz offset, for the same sideband offset frequency, electron-hole pairs driven by higher-frequency fields take significantly longer to recollide, because, outside the linear region, the higher-frequency fields are significantly weaker than the lower-frequency fields. The longer time increases the chance for the particles to scatter, lowering sideband powers and comb bandwidths, which matches the observed trends in the top left of Fig.~\ref{fig:5bandwidthComps}.

These trends are also evident in the other panels of Fig.~\ref{fig:5bandwidthComps}, with three major exceptions. For the 5nm QW at low field (lower left Fig.~\ref{fig:5bandwidthComps}) the comb produced from the 540 GHz field was stronger than those produced at lower frequencies. For the higher fields (Fig.~\ref{fig:5bandwidthComps} right), the 340 GHz combs were significantly stronger and wider for both samples--even at sideband offsets smaller than the 7 THz sideband offset that corresponds to a 10\% deviation of the 640 GHz electric field from the linear approximation.  

While the proposed scaling law, with adjustments to account for deviations from the linear approximation, captures many observed trends, it is clearly insufficient to describe all of the data.  The assumption that electrons and holes each travel in a single, parabolic band may be one of the problems. In GaAs QWs, there are several closely-spaced and strongly-mixed valence subbands, with several low-energy avoided crossings. It has been shown that sideband polarizations and intensities are sensitive to the motion of holes through these avoided crossings ~\cite{DBR}. Accounting for the influence of the complexity of the valence subbands on the scaling of HSG is outside the scope of this work.

\section{Measuring the free electron laser}
As an application of OFCs from HSG, we use them to characterize the Free Electron Laser (FEL) used in these experiments. The THz regime lacks the variety of well-developed methods for measuring the frequency or linewidth of a source over a large range that is enjoyed in the optical and infrared frequency ranges. Often, specialized setups are designed for specific applications. For example, the UCSB high-field electron paramagnetic resonance spectrometer makes use of a heterodyne mixing scheme~\cite{susumuIL, susumuCDC}, but is limited to a narrow bandwidth; electro-optic sampling (EOS) is a phase-sensitive technique and widely used for performing THz spectroscopy~\cite{EOSfirst}, but requires phase stable source to average over many pulses; and holographic Fourier transform techniques have been demonstrated~\cite{Agladze2010}, but require high power and expensive array detectors. HSG, however, is a technique for measuring the frequency and, as we will show, linewidth of the driving THz field over a large frequency range, on a phase-unstable source, while being robust to alignment. 

The frequency of THz FELs have been measured using first-order, perturbative sideband generation based on the linear electro-optic effect, in which $f_\text{SB} = f_\text{NIR} \pm f_\text{THz}$. Sidebands were detected using apparatus similar to that used in this work~\cite{singleshotspectrometer,Wijnen:10}. In these measurements, if the uncertainty in the NIR laser frequency is negligible, the uncertainty in the THz frequency $\delta f_\text{THz}$ is limited by the resolution of the sideband frequency $\delta f_\text{SB}$, which is determined by the monochromator and CCD. In an HSG spectrum, the sidebands are seen at $f_\text{SB} = f_\text{NIR} + nf_\text{THz}$ ($n$ even), and the corresponding uncertainty of the THz frequency, $\delta f_\text{THz}$, is $\frac{1}{n}[\delta f_\text{NIR}^2 + \delta f_\text{SB}^2]^{1/2}$. If the frequency of the NIR laser is measured or known to high precision, $\delta f_\text{NIR}\ll\delta f_\text{SB}$, the uncertainty in the THz frequency is then approximately $\frac{1}{n}\delta f_\text{SB}$, more than 100 times smaller than $\delta f_\text{SB}$ if $n>100$. Each of the frequencies reported in Fig.~\ref{fig:2rawcombs} comes from the slope of the linear fit of sideband frequency versus sideband order. The covariance and residuals from the fits indicate the uncertainty of the frequency is within 1 GHz.

\begin{figure}
    \centering
    \includegraphics{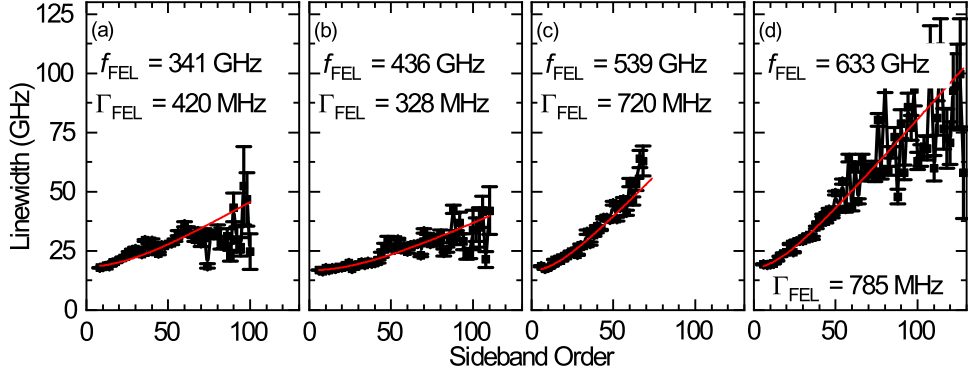}
    \caption{Linewidths of sidebands taken as the Gaussian widths from the fit of the spectra shown in Fig.~\ref{fig:2rawcombs}. Each spectra is fit to Eq.~\ref{eq:broadening} to extract the linewidth of the FEL at each frequency. }
    \label{fig:3Linewidths}
\end{figure}

The linewidth of the UCSB FEL output is too small to measure using first-order, perturbative sideband generation. However, it can be measured using HSG.  
The linewidth of each sideband, $\Gamma_\text{SB}$, depends on the resolution of the spectrometer, $\Gamma_\text{spec}$, as well as the linewidth of the FEL, $\Gamma_\text{FEL}$. A THz pulse from the UCSB FEL contains contributions from multiple longitudinal modes of the cavity, resulting in each pulse containing several of these modes simultaneously. Competition between modes causes pulse-to-pulse fluctuations in the spectral contents of each pulse~\cite{susumuIL}, in addition to fluctuations in radiated power. This inhomogenous broadening results in a sideband linewidth given by

\begin{equation}
    \Gamma_\text{SB} = \sqrt{\Gamma_\text{spec}^2 + n^2\Gamma_\text{FEL}^2}\label{eq:broadening}
\end{equation}

Figure~\ref{fig:3Linewidths} shows the linewidths of the sidebands from the combs shown in Fig.~\ref{fig:2rawcombs}, and the corresponding fits to Eq.~\ref{eq:broadening}. At low orders, $n\lesssim 20$, the linewidths are determined by the resolution of the spectrometer, which is approximately 20 GHz, depending on experimental conditions. The FEL linewidth contributes to the measured linewidth of sidebands only above about 20th order. From Fig.~\ref{fig:3Linewidths}, the extracted FEL linewidths are $\Gamma_\text{FEL, 341 GHz}=420\pm21$ MHz, $\Gamma_\text{FEL, 436 GHz}=328\pm12$ MHz, $\Gamma_\text{FEL, 539 GHz}=720\pm16$ MHz, and $\Gamma_\text{FEL, 633 GHz}=785\pm15$ MHz with errors taken as the covariance of the fits to Eq.~\ref{eq:broadening}. At 247 GHz, the low field strength and low frequency resulted in too few sidebands observed to accurately determine the FEL linewidth. At the remaining THz frequencies, the FEL linewidth is well resolved and below 1 GHz, matching previous measurements with a sub-band harmonic mixer, which only works in a very small frequency range~\cite{susumuIL}.

\section{Conclusion}
High-order sideband generation provides a mechanism for generating optical frequency combs with bandwidths exceeding 80 THz with continuous control of the offset frequency and comb spacing. Our results demonstrate the ability to tune the comb offset frequency throughout the comb spacing by varying the NIR laser frequency near the band gap of the GaAs QWs used. Combs centered at different frequencies can be created by choosing materials with band gaps in the desired frequency range, for example by choosing quantum wells with different widths or compositions~\cite{DBR, Zaks2012}, or using altogether different semiconductors~\cite{Langer2016, bulkGaAs}. Furthermore, Fig.~\ref{fig:2rawcombs} shows the ability to tune the comb spacing between 480 GHz to 1280 GHz. Much higher spacings of 3 THz~\cite{Zaks2012} or even 60 THz~\cite{Langer2016} are possible. We have shown that OFCs based on HSG are practical for measuring the linewidth of our THz source over a large frequency range. While the high power of an FEL was utilized in experiments here, careful microwave engineering~\cite{Zaks2012} or use of antenna structures~\cite{banksAntenna} can provide comparable fields using compact sources with much lower power than the FEL. Demonstration of HSG using such a compact source would open the door to applications of OFCs based on HSG in metrology, spectroscopy, telecommunications, and other areas. The results of this paper illustrate some of the tradeoffs between THz field and frequency that will be required in the design of OFCs based on HSG.

\section{Acknowledgements}

The work performed at UCSB was done with the support of grant NSF DMR 1405964. The work at Princeton University was funded by the Gordon and Betty Moore Foundation through the EPiQS initiative Grant GBMF4420, and by the National Science Foundation MRSEC Grant DMR 1420541. Work performed at the U.S. Naval Research Laboratory was supported by the Office of Naval Research. 




\end{document}